\documentclass[preprint,nofootinbib,longbibliography]{revtex4-1}
\usepackage[utf8]{inputenc}
\usepackage{amsmath}
\usepackage{amsfonts}
\usepackage{xcolor}

\usepackage[colorlinks=true]{hyperref} 
\hypersetup{
    bookmarks=true,         % show bookmarks bar?
    unicode=false,          % non-Latin characters 
    pdftoolbar=true,        % show Acrobat
    pdfmenubar=true,        % show Acrobat 
    pdffitwindow=false,     % window fit to page when opened
    pdfstartview={FitH},    % fits the width of the page to the window
    pdfsubject={},   % subject of the document
    pdfcreator={},   % creator of the document
    pdfproducer={}, % producer of the document
    pdfkeywords={} {} {}, % list of keywords
    pdfnewwindow=true,      % links in new window
    colorlinks=true,       % false: boxed links; true: colored links
    linkcolor=magenta, %red,          % color of internal links (change box color with linkbordercolor)
    citecolor=blue,        % color of links to bibliography
    filecolor=magenta,      % color of file links
    urlcolor=blue           % color of external links
} 

\newcommand{\bx}{{\mathbf x}}
\newcommand{\by}{{\mathbf y}}
\newcommand{\bb}{{\mathbf b}}
\newcommand{\bk}{{\mathbf k}}

\begin{document}

\title{Hohenberg-Mermin-Wagner-type theorems and dipole symmetry}
\author{Anton Kapustin}
\email{kapustin@theory.caltech.edu}
\affiliation{California Institute of Technology, Pasadena, CA 91125, United States}
\author{Lev Spodyneiko}
\email{lionspo@mit.edu}
\affiliation{Massachusetts Institute of Technology, Cambridge, MA 02139, United States}

\begin{abstract}
We study the possibility of spontaneous symmetry breaking in systems with both charge and dipole symmetries. For $d$-dimensional systems at a positive temperature, we show that charge symmetry cannot be spontaneously broken for $d\leq 4$, while dipole symmetry cannot be spontaneously broken for $d\leq 2$. For $T=0$, we show that charge symmetry cannot be spontaneously broken for $d\leq 2$ if the compressibility is finite. We also show that continuum systems with a dipole symmetry have infinite inertial mass density.
\end{abstract}
\maketitle
\section{Introduction}\label{sec: intro}

The concepts of spontaneous symmetry breaking and long-range order play a central role in the theory of phase transitions. R. E. Peierls \cite{PeierlsG,PeierlsF} and L. D. Landau \cite{Landau} were the first to realize that in sufficiently low dimensions thermal fluctuations of the order parameter can preclude genuine long-range order. The impossibility of long-range order in one and two dimensions at positive temperatures has been proved by P. C. Hohenberg \cite{Hohenberg} in the case of superfluids and superconductors and by N. D. Mermin and H. Wagner \cite{MW} for (anti)ferromagnets. The relevant symmetries in these cases are either $U(1)$ or $SU(2)$. 
Analogous theorems for one-dimensional systems at zero temperature were proved by L. Pitaevskii and S. Stringari \cite{Pitaevskii1991}. At $T=0$ it is quantum fluctuations that destroy the order.

The impossibility of symmetry breaking in a particular system does not guarantee the absence of continuous phase transitions. A textbook example  is the BKT phase transition \cite{Berezinski1,Berezinski2,Kosterlitz}. In this case fluctuations of the order parameter are qualitatively different on the two sides of the transition, even though symmetry is not broken. 

The main purpose of this paper is to generalize the above results to $U(1)$-invariant systems which conserve not only the charge but also the dipole moment. The additional symmetry possessed by such models is called the dipole symmetry. A simple non-interacting field theory with a dipole symmetry is the Lifshitz model which has a single real scalar field $\phi(t,\bx)$ and an action
\begin{equation}\label{eq:lifshitz}
S=\frac12 \int dt\, d^dx \left({\dot \phi}^2- C (\partial^2\phi)^2\right).
\end{equation}
Here $C$ is a positive constant and $\partial^2 = \partial_j \partial_j$. 
This model has both a charge symmetry $\phi\mapsto\phi+a$ and a dipole symmetry $\phi\mapsto \phi+\bb\cdot \bx$, where $\bb$ is a constant vector.
An example of a lattice system with a dipole symmetry dubbed the dipolar Bose-Hubbard model was recently studied by E.~Lake et al. \cite{lakesenthil}. Its Hamiltonian is given by
\begin{align} \label{eq: bose hubbard}
\begin{split}
    H=  \sum_{i}&\Bigg( -t\sum_{a}\left[{b^\dagger_i}^2 b_{i+a}b_{i-a}+b_{i-a}^\dagger b_{i+a}^\dagger {b_i}^2 \right]\\&-t'\sum_{a,b,b\ne a}\left[ b^\dagger_ib^\dagger_{i+a+b} b_{i+a}b_{i+b} +b^\dagger_{i+b}b^\dagger_{i+a}b_{i+a+b}b_i \right]+ \frac U 2n_i(n_i-1)-\mu n_i \Bigg),
\end{split}
\end{align}
where $b_i$ are boson creation operators and $n_i = b^\dagger_i b_i$, $i$ goes over square lattice sites and $a,b$ over basis vectors. The dipole symmetries are generated by $\sum_j j_a n_j$ where $j_a$ is the $a$-coordinate of the site $j$. The usual hoping term is forbidden by the dipole symmetry and particles can only hop in pairs. Ref. \cite{lakesenthil} shows that within the mean-field approximation the model has several different phases characterized by spontaneous breaking of dipole and/or charge symmetries. 

Our interest in such models stems from a connection between dipole symmetry and systems where the kinetic energy of charged quasi-particles is suppressed relative to the potential interaction energy, leading to strongly correlated physics. Whenever a system is effectively described by weakly interacting quasi-particles, suppression of the kinetic energy is tantamount to the flatness of the relevant band. For systems with strongly interacting quasi-particles, the concept of a flat band is problematic. A possible way out is to notice that an idealized exactly flat band has an additional symmetry of shifts by a constant momentum, or equivalently a Galilean boost symmetry. Invariance of the Hamiltonian under boost can be used as a definition of a ``flat band'' which does not rely on the weakness of the interaction.  Although in the experimentally realized systems the band is never exactly flat (i.e. the commutator of the boost generator and the Hamiltonian is never exactly zero), the study of such idealized models is a good starting point to tackle the theoretical challenges of strong interactions in these materials.

To explain the  connection to the dipole symmetry, consider a system of particles with mass $m$ and density-density interactions. The generator of a shift in momentum space is given by
\begin{align}\label{eq: boost}
    {\bf K}=\int d^dx  \left(t\frac{{\bf p}({\bf x})}m - {\bf x} \rho({\bf x})\right)
\end{align}
where ${\bf p}({\bf x}),\rho({\bf x})$ are the momentum and particle densities and we divided the boost generator by $m$ since we want it to shift momentum instead of velocity. In the limit $m\rightarrow \infty$ the kinetic energy become quenched and ${\bf K}$ starts to commute with the Hamiltonian, i.e. becomes a true symmetry instead of being part of the spectrum-generating algebra. In this limit, the generator (\ref{eq: boost}) becomes the total dipole moment of the particles.

Dipole moment conservation on top of  charge conservation leads to restricted mobility of charged particles but does not obstruct the propagation of neutral bound states. Similar phenomena appear in theories of fractons \cite{Haah,Fu,FractonReview} and their connection to dipole symmetry is known \cite{Pretko,Pretko2,Seiberg}. Therefore models with dipole symmetries and fractonic behavior can be used as a  playground for studying  systems with vanishing kinetic energy. One can hope that their exotic features can also manifest themselves in more realistic ``flat-band'' systems. 

The issue of spontaneous breaking of dipole symmetry is related to the question of whether systems that are invariant under boost symmetry in the above sense can nevertheless have quasi-particles with a nonzero dispersion. To see that this might be possible, consider the famous Landau relation between the bare mass $m$ and the quasi-particle mass $m^*$ in a Fermi liquid \cite{AGD}:
\begin{align}
    \frac{1}{m^*} = \frac 1  m  - \frac{2p_f}{(2\pi)^3} \int  d \Omega \, f(\theta) \cos \theta.
\end{align}
This relation follows from Galilean invariance. One can see that in the limit $m\rightarrow \infty$ (and assuming that the system stays in the Fermi liquid phase) the effective mass of quasiparticles can be generated by the interaction and remain finite. Since the Fermi surface is not invariant under translations in momentum space, for this mechanism to work the dipole symmetry of the  $m=\infty$ model must be spontaneously broken. 

Within the Lifshitz model, the effects of quantum and thermal fluctuations on the existence of phases with broken charge or dipole symmetries have been discussed in \cite{FractonSuperfluid,Lake2} and are  reviewed in Section \ref{sec: Lifshitz}.
This type of analysis assumes that the only low-energy degrees of freedom are Goldstone bosons. However, spontaneous breaking of dipole symmetry (or more generally any symmetry that has a non-trivial commutator with translations) does not necessarily lead to a Goldstone mode \cite{Watanabe,Alberte}. For example,  some Fermi liquids  spontaneously break rotational and/or Galilean symmetries but their low energy excitation spectrum consists of the particle-hole continuum without any bosonic  collective modes. A further complication is that the choice of the fields describing the  Goldstone modes is not unique. For example, in the traditional approach to crystalline solids, one does not introduce a separate field for the rotational symmetry, even though both rotational and translational symmetries are broken to a discrete subgroup. But the same low-energy physics can also be described by a field theory with additional fields which play the role of the Goldstone bosons for rotations \cite{LeoMike,Leo}.
Thus it is desirable to prove theorems of  Hohenberg-Mermin-Wagner (HMW) type  without assuming anything about the low-energy degrees of freedom.

In Section \ref{sec: lattice}, we consider a general $d$-dimensional lattice system with a finite range Hamiltonian that commutes with both charge and dipole moment. We show that for $T>0$ the dipole symmetry cannot be broken for $d\le2$,  while the charge symmetry cannot be broken for $d\le4$. Thus the mere presence of dipole symmetry allows one to strengthen the conclusions of the HMW theorem. At $T=0$  we show that the charge symmetry cannot be broken for $d\le2$ if the charge compressibility does not diverge. The theorem does not put any restriction on the dipole symmetry at zero temperature because the dielectric constant diverges for a general system with a dipole symmetry. Our results put on firm ground the analysis of \cite{FractonSuperfluid,Lake2} based on the Lifshitz model.

In Section \ref{sec: continuum} we prove HMW-type  theorems for continuum models with a dipole symmetry. Unlike lattice systems, such models  have a conserved momentum ${\mathbf P}$, and thus one may also consider a generalized Gibbs ensemble with a nonzero velocity ${\mathbf v}$.
%(i.e. the ensemble where the Hamiltonian $H$ is replaced with $H-{\mathbf v}\cdot{\mathbf P}$). 
We show that the inertial mass density in continuum models with a dipole symmetry is always infinite. This is a kind of converse to the statement that dipole symmetry emerges as the infinite-mass limit of the Galilean symmetry. Section \ref{sec: discussion} contains concluding remarks.

We thank  Xiaoyang Huang, Ethan Lake, Leo Radzihovsky, and Senthil Todadri for discussions. AK was supported in part by the U.S.\ Department of Energy, Office of Science, Office of High Energy Physics, under Award Number DE-SC0011632. AK was also supported by the Simons Investigator Award. LS was supported by the Simons Collaboration on Ultra-Quantum Matter, which is a grant  (651446) from the Simons Foundation.

\section{The Lifshitz model}\label{sec: Lifshitz}

The Lifshitz model (\ref{eq:lifshitz}) in $d$ spatial dimensions is the simplest field theory which  conserves both charge and its dipole moment. The charge symmetry is generated by a Hermitian operator $Q$ satisfying $[Q,\phi]=-i$. The dipole symmetry is generated by a vector-valued Hermitian operator ${\bf D}$ satisfying $[{\bf D},\phi(t,\bx)]=-i\bx$. Naively, the Lifshitz model describes a phase where  both charge and dipole symmetries are completely broken, with $\phi$ being the order parameter. Indeed, if $\langle\cdot\rangle$ denotes averaging over a ground state or a thermal state, then we have $\langle [Q,\phi(t,\bx)]\rangle=-i\neq 0$ and $\langle [{\bf D},\phi(t,\bx)]\rangle=-i\bx\neq 0$. By definition, this means that symmetries generated by both $Q$ and ${\bf D}$ are spontaneously broken. But it is well known that quantum and thermal fluctuations may invalidate this conclusion by making $\phi$ ill-defined. Let us examine this issue in more detail.

For $T=0$ and $d>2$ the connected two-point function of $\phi$ is 
\begin{equation}
\langle \phi(0,\bx)\phi(0,\by)\rangle_c\sim \frac{1}{|\bx-\by|^{d-2}}.
\end{equation}
This is a well-defined two-point function that satisfies the cluster decomposition. Thus $\phi$ is a well-defined local field that may serve as an order parameter for both charge and dipole symmetry breaking. In contrast, for $d=2$ the connected two-point function of $\phi$ is not well-defined, because its spatial Fourier transform
\begin{equation}\label{eq:sing}
\langle \phi(0)_{\bk}\phi(0)_{\bk}\rangle_c\sim \frac{1}{k^2}
\end{equation}
has a non-integrable singularity at $\bk=0$.\footnote{If one cuts off the integral at $k=\Lambda_{\rm IR}$, one finds a logarithmically growing two-point function.} Less formally, the average of $\phi(0,\bx)$ over a large region of diameter $L$ has a standard deviation that grows like $(\log L)^{1/2}$, indicating that it cannot serve as an order parameter. 

On the other hand, the two-point function of the vector-valued operator ${\bf u}=\nabla\phi$ has an integrable singularity in momentum space, and the two-point function in coordinate space is well-defined and satisfies the cluster decomposition:
\begin{equation}
\langle u_j(0,\bx) u_l(0,\by)\rangle_c\sim \frac{(x_j-y_j)(x_l-y_l)}{|\bx-\by|^4}
\end{equation}
Since $\bf u$ commutes with $Q$, it cannot serve as an order parameter for charge symmetry breaking. Neither there are any other local fields whose commutator with $Q$ has a non-zero expectation value. Thus charge symmetry is unbroken in the $d=2$ Lifshitz model at zero temperature. But since $\langle [D_j,u_l]\rangle=-i\delta_{jl}$, the  dipole symmetry is spontaneously broken. 

The situation for $d=1$ is similar. The two-point function of the field $\phi$ in momentum space again has a non-integrable singularity (\ref{eq:sing}). The standard deviation of the spatial average of $\phi$ grows like $\sqrt L$ for large $L$. Since one lacks any order parameter for the charge symmetry breaking, it remains unbroken. But the two-point function of the field $u=\partial_x\phi$ is constant in momentum space and thus we have
\begin{equation}
\langle u(0,x) u(0,y)\rangle_c\sim \delta(x-y). 
\end{equation}
This is a well-defined (although somewhat unusual) answer, and $u$ can serve as an order parameter for the dipole symmetry breaking. Thus the dipole symmetry in the $d=1$ Lifshitz model is broken at $T=0$, as noted in \cite{Lake2}.\footnote{Spontaneous breaking of a continuous symmetry in a 1d system might seem surprising, but it does not contradict any known general theorems. Another example is a system of non-relativistic fermions in 1d which spontaneously breaks boost invariance by forming a Luttinger liquid.}

For $T>0$ the static two-point function of the Lifshitz field $\phi$ in momentum space is
\begin{equation}\label{eq:singT}
    \langle \phi(0)_{\bk}\phi(0)_{\bk}\rangle_c\sim \frac{1}{k^4}
\end{equation}
For $d>4$ it is integrable, in which case the connected thermal two-point function of $\phi$ in coordinate space is well-defined. Since $\phi$ is well-defined and shifts under charge symmetry, the charge symmetry (and therefore also the dipole symmetry) are spontaneously broken for $d>4$. For $d\leq 4$, the field $\phi$ is not well-defined, and there are no other order parameters for charge symmetry breaking. Thus the charge symmetry is not broken for $d\leq 4$. As for dipole symmetry, for $d>2$ it is spontaneously broken, while for $d\leq 2$ it is also unbroken. Indeed, for $d> 2$ we can take $u_j=\partial_j\phi$ as an order parameter, since its connected thermal two-point function has an integrable singularity $\sim 1/k^2$ in momentum space and decays as $1/x^{d-2}$ in coordinate space. 

%On the basis of this example, one may advance several conjectures about the properties of general quantum systems with dipole symmetries.
%\begin{itemize}
    %\item For $T>0$ charge symmetry is unbroken whenever $d\leq 4$.
    %\item For $T>0$ non-constant dipole symmetry is unbroken whenever $d\leq 2$.
    %\item For $T=0$ charge symmetry is unbroken whenever $d\leq 2$ and compressibility is finite. 
%\end{itemize}
%The condition of finite compressibility is imposed by analogy with \cite{Pitaevskii1991}.
%Below we prove these conjectures for quantum lattice systems with finite-range interactions. For $T>0$ we adapt the arguments of Hohenberg \cite{} and Mermin-Wagner \cite{}, while for $T=0$ we modify the arguments in \cite{}.

\section{Lattice systems}\label{sec: lattice}

\subsection{Definitions}
We will consider a lattice system with discrete translation symmetry and finite-range interactions. More formally, let $\Lambda$ be a lattice in $\mathbb R^d$ with an action of translation group $\mathbb Z^d$. The Hilbert space is the product of an infinite number of copies of a finite-dimensional Hilbert space of a single site. Local operators are operators which act non-trivially only on a finite number of sites. We can formally combine all sites in a single unit cell into a single site and reduce the lattice $\Lambda$ to $\mathbb Z^d$ without a loss of generality.  If the translational symmetry is broken to a sublattice symmetry (as in antiferromagnets), we can formally combine the resulting unit cell to one cell and again assume that the lattice is $\mathbb Z^d$.

The Hamiltonian density operator $H_{\bf n}$ is assumed to be translationally invariant and finite-range. The integer-valued index ${\bf n}$ enumerates the lattice $\mathbb Z^d$ sites.  Moreover, we assume that the system has a charge symmetry and an associated dipole symmetry. Namely, there exist charge density operators $Q_{\bf n}$ which are finite-range, translationally invariant, and satisfy
\begin{align}
    [Q_{\bf n},Q_{\bf m}]&=0,\\
    \sum_{{\bf n}\in \mathbb Z^d} [Q_{\bf n}, H_{\bf m}]&=0,\\
    \sum_{{\bf n}\in \mathbb Z^d} n_i[Q_{\bf n}, H_{\bf m}]&=0, \quad \text{for all $i=1,\dots, d$},
\end{align}
where $n_i$ is $i$-th component of the ${\bf n}$ in $\mathbb Z^d$. The latter condition means the that dipole moment is conserved. We will define the dipole charge density as $D^i_{\bf n} = n_i Q_{\bf n}$ and the dipole charge as $D^i = \sum_{{\bf n} \in \mathbb Z^d} D_{\bf n}^i$. The operators $Q_{\bf n}, H_{\bf n}$ are assumed to be local operators supported on balls of diameter $d_{\rm int}$ centered at ${\bf n}$. 

% If the charge density is charged itself, i.e.
% \begin{align}
%     [Q,Q_n] \ne 0,
% \end{align}
% then we can separate it into a family of charges with a fixed charge with respect to $Q$. Each of these will be conserved separately. Therefore, by subtracting charged part we can always redefine the charge to be neutral with respect to itself
% \begin{align}
%     [Q,Q_n] = 0.
% \end{align}
% The same choice can be made for $D^i$
% \begin{align}
%     [D^i,D^i_n]=0.
% \end{align}

We also assume that the state of the system has a clustering property. More precisely, following \cite{Martin1982,Momoi1996}, we assume that for any two local operators $A,B$ one has
\begin{align}
    |\langle A T_{\bf n}(B)\rangle- \langle A\rangle \langle B\rangle | \le \frac{C_{AB}} {|{\bf n}|^{\delta_{AB}} }\quad \text{as} \quad |{\bf n}|\rightarrow \infty,
\end{align}
where $T_{\bf n}$ is the translation operator and $\delta_{AB},C_{AB}$ are positive numbers. In reality, we only use the clustering property when $A=B$ and $A$ is a local operator which is a candidate for the order parameter. Some sort of clustering is also implicit in \cite{Hohenberg,MW}, since it is assumed there that the Fourier transform of a two-point correlator is a well-defined integrable function rather than a distribution.

\subsection{HMW theorem for charge symmetry at a non-zero temperature}

We repeat the argument of \cite{Momoi1996,Martin1982} making only small adjustments. Suppose the charge symmetry is spontaneously broken, i.e. there exists a local operator $A_0$ such that 
\begin{equation}\label{eq: broken charge T ne 0}
    \langle [Q, A_0]\rangle \ne 0, 
\end{equation}
where brackets denote the thermal average. Without a loss of generality, we may assume that $A_0$ is Hermitian, has zero average $\langle A_0\rangle=0$ and is localized around the point ${\bf n} = 0$. 

We will consider a spatially-averaged version of $A_0$ defined as
\begin{align}\label{eq: A definiton}
    A_R = \frac 1 {{\rm Vol}(B^d_R)}\sum_{|{\bf n}|\le R} T_{\bf n} (A_0),
\end{align}
where $T_{\bf n}$ is the translation operator, $R$ is some number, and ${\rm Vol}(B^d_R)$ is the number of lattice sites in a $d$-dimensional ball of radius $R$. 
We assume that the full $\mathbb Z^d$ translational symmetry is preserved and thus 
\begin{equation}\label{A0AR}
\langle [Q, A_0]\rangle=\langle [Q,A]\rangle .
\end{equation}
By increasing $d_{\rm int}$ if necessary, we may assume that $A_0$ is supported on a ball of diameter $d_{\rm int}$ with a center at $0$. Then $A_R$ is supported on a ball of diameter $2R+d_{\rm int}$ centered at $0$. 
%We can act with a translation operator on the left side of (\ref{eq: broken charge T ne 0}) without affecting the right-hand side. 
 
Without changing the commutator, we can also replace $Q$ in eq. (\ref{A0AR}) with $Q(f) = \sum_{{\bf n} \in \mathbb Z^d} f({\bf n}) Q_{\bf n}$, where $f({\bf n})$ is any compactly supported function equal to $1$ whenever ${\bf n}$ belongs to a ball of diameter $2R+3d_{\rm int}$ centered at $0$. 

We will use the Bogoliubov inequality valid for any two bounded operators $A,B$ and a thermal (Kubo-Martin-Schwinger) state at a nonzero temperature $T$:
\begin{align} \label{eq: Bog ineq}
    |\langle [A,B]\rangle |^2 \le \frac 1 {2T} \langle [B,[H,B^\dagger]]\rangle  \langle A^\dagger A +A A^\dagger\rangle .
\end{align}
We will apply it to $A=A_R$ and $B=Q(f)$.

\subsubsection{Estimates} \label{sec: Q estimates}

We want to estimate $\langle[Q(f),[Q(f),H]]\rangle$. We can rewrite the commutators as
\begin{align}\label{eq: Q double com}
\begin{split}
    [Q(f),H] &= \sum_{{\bf m},{\bf k}} f({\bf m}) [Q_{\bf m},H_{\bf k}] = \sum_{{\bf m},{\bf k}} (f({\bf m})-f({\bf k})) [Q_{\bf m},H_{\bf k}],  \\
    [Q(f),[Q(f),H]] &=\sum_{{\bf n},{\bf m},{\bf k}} f({\bf n})(f({\bf m})-f({\bf k}))  [Q_{\bf n}, [Q_{\bf m},H_{\bf k}]] \\&= \sum_{{\bf n},{\bf m},{\bf k}} (f({\bf n})-f({\bf k}))(f({\bf m})-f({\bf k})) [Q_{\bf n}, [Q_{\bf m},H_{\bf k}]], 
\end{split} 
\end{align}
The commutator $\langle[Q_{\bf n}, [Q_{\bf m},H_{\bf k}]]\rangle$ is non-zero only if the supports of $H_{\bf k}$, $Q_{\bf n}$ and $Q_{\bf m}$ overlap. The expectation value can be estimated as
\begin{align}\label{eq: double comm estimate}
    \left|\langle[Q_{\bf n}, [Q_{\bf m},H_{\bf k}]]\rangle\right| \le 4 \|Q_0\|^2 \|H_0\|,  
\end{align}
where the operator norm is defined as $\|A\|=\max_{v\in \mathcal H} \frac {|Av|}{|v|}$.

We can expand the function $f$ in the Taylor series around $\bf k$ with the reminder in the Lagrange form
 \begin{align}
     f({\bf n}) = f({\bf k})+(({\bf n}-{\bf k})\cdot \nabla) f({\bf k}) +\frac 1 2 (({\bf n}-{\bf k})\cdot \nabla)^2 f(\xi),
 \end{align}
 where $({\bf n}-{\bf k})\cdot \nabla$ is derivative in the direction of the vector ${\bf n}-{\bf k}$ and $\xi$ is point on the straight interval from ${\bf k}$ to ${\bf n}$. Due to dipole symmetry linear terms proportional to $(({\bf n}-{\bf k})\cdot \nabla) f({\bf k})$ and $(({\bf m}-{\bf k})\cdot \nabla) f({\bf k})$ in (\ref{eq: Q double com}) will be zero and we can estimate  
  \begin{align} \label{eq: QQ com estimate}
 \left|\langle[Q(f),[Q(f),H]]\rangle \right|\le C_d R_f^d d_{\rm int}^4 \|Q_0\|^2 \|H_0\| \big(\max |\nabla^2f|   \big)^2,
 \end{align}
 where $R_f$ is half of the diameter of the support of the function $f$ and $C_d$ is a constant that depends only on $d$.
 
 Choosing the function $f({\bf r})$ to be $1$ for $r\leq R+\frac{3}{2} d_{\rm int}$, $0$ for $r>2R$ and smoothly interpolating between these two values with the second derivatives of order $O(\frac 1 {R^2})$, we find
 \begin{align} \label{eq: q double com estimate}
 \langle[Q(f),[Q(f),H]]\rangle = O(R^{d-4}).
 \end{align}

Using the clustering property and letting $\delta=\delta_{A_0 A_0}$, we can estimate

\begin{align}\label{eq: clustering estimate}
\begin{split}
    \langle A_R^\dagger A_R +A_R A_R^\dagger\rangle = \frac{2}{{{\rm Vol}(B^d_R)}^2} \sum_{|{\bf n}|\le R,|{\bf m}|\le R}\langle T_{\bf n}(A_0) T_{\bf m}(A_0) \rangle\\<  \frac{2C_{A_0A_0}}{{{\rm Vol}(B^d_R)}} \sum_{|{\bf n}|<2R} \frac 1 {|{\bf n}|^\delta} =
    \begin{cases}
        O\left(\frac 1 {R^\delta}\right) &\text{if} \quad \delta<d,\\
            O\left(\frac {\ln R} {R^d}\right) &\text{if} \quad \delta=d,\\
                O\left(\frac 1 {R^d}\right) &\text{if} \quad \delta>d.
    \end{cases}
\end{split}
\end{align}

Combining the two estimates and substituting them into the Bogoliubov inequality we find
 \begin{align}\label{eq: HMW charge}
    | \langle [Q, A_0]\rangle|^2  = |\langle [Q(f), A_R]\rangle|^2 =  O(R^{d-4-\delta}).
 \end{align}
 
Since $R$ is arbitrary, this means that for $d\leq 4$ and any local $A_0$ we have $\langle [Q,A_0]\rangle =0$. Thus the charge symmetry cannot be spontaneously broken for $d\le 4$ if the temperature is non-zero.

\subsection{HMW theorem for the dipole symmetry at a non-zero temperature}\label{sec:dipole}
We can use the estimates from the previous section to constrain dipole symmetry breaking. 

Suppose dipole symmetry is spontaneously broken, with a local order parameter $A_0$:
\begin{equation}\label{eq: broken dipole T ne 0}
    \langle [D^i, A_0]\rangle \ne 0.
\end{equation}
We will assume that the charge symmetry is not broken\footnote{If the charge symmetry is broken then the dipole symmetry is broken as well.}. This will allow us to act with a translation operator on the left-hand side of (\ref{eq: broken dipole T ne 0}) without affecting the right-hand side. We will consider a spatially-averaged $A_0$ given by (\ref{eq: A definiton}). We can replace $D^i$ in this equation with $D^i(g) = \sum_{{\bf n} \in \mathbb Z^d} g({\bf n}) D^i_{\bf n}$, where $g({\bf n})$ is any compactly supported function equal to $1$ whenever ${\bf n}$ belongs to a ball of diameter $2R+3d_{\rm int}$ centered at $0$.

We will use the Bogoliubov inequality (\ref{eq: Bog ineq}) with $A=A_R$ and $B= D^i(g)$.

\subsubsection{Estimates}

We now want to estimate 
\begin{align*}
    \langle[D^i(g),[D^i(g),H]]\rangle &=  \sum_{{\bf n},{\bf m}}g({\bf n}) g({\bf m})\langle  [D_{\bf n}^i,[D^i_{\bf m},H]]\rangle\\ &= \sum_{{\bf n},{\bf m}} n^i g({\bf n}) m^i g({\bf m})\langle [Q_{\bf n},[Q_{\bf m},H]]\rangle.
\end{align*}
The considerations of the previous section did not use a specific form of the function $f$.  Therefore, we can  use the same estimate (\ref{eq: QQ com estimate})  with the  function $f({\bf n})=n^i g({\bf n})$:
\begin{align} \label{eq: DD com estimate}
 |\langle[D^i(g),[D^i(g),H]]\rangle | \le C_d(R_f)^d d_{\rm int}^4 \|Q_0\|^2 \|H_0\| \big(\max |\nabla^2f|   \big)^2.
 \end{align}

Since $g({\bf r})$ must be $1$ for $r\leq R+\frac{3}{2}d_{\rm int}$ the function $f({\bf r})$ must be equal to $x^i$ for $r\leq R+\frac{3}{2}d_{\rm int}$. We can smoothly interpolate it to zero for $r>2R$ while keeping the second derivatives bounded by $O(\frac 1 {R})$. Then we get 
\begin{align}
    |\langle[D^i(f), [D^i(f),H]]\rangle | = O(R^{d-2}).
\end{align}

 Using this estimate, the Bogoliubov inequality, and the estimate (\ref{eq: clustering estimate}), we find
  \begin{align}\label{eq: HMW dipole}
     |\langle [D^i, A_0]\rangle|^2  = |\langle [D^i(f), A]\rangle|^2 =  O(R^{d-2-\delta}).
 \end{align}
Since $R$ is arbitrary, the dipole symmetry cannot be spontaneously broken in $d\le2$ at non-zero temperatures.

 \subsection{HMW-type theorems at zero temperature}
 
 At zero temperatures instead of the Bogoliubov inequality we use the uncertainty relation \cite{Pitaevskii1991}:
 \begin{align}\label{eq: uncertainty relation}
     |\langle [A,B]\rangle|^2 \le \langle A^\dagger A +AA^\dagger\rangle \langle B^\dagger B +BB^\dagger\rangle ,
 \end{align}
where brackets denote the average over any state and we assumed that the operators $A,B$ have zero averages. We will apply it to the case when $A=A_R-\langle A_R \rangle$, $B=Q(f)-\langle Q(f)\rangle$, and the state is the ground state of $H$. We can upper-bound the right-hand side of this inequality in terms of more physical quantities as follows. Recall that for any two local operators $B,B'$ one  defines the structure factor $S_{BB'}(\omega)$ as 
\begin{equation}
S_{BB'}(\omega)=\int_{-\infty}^\infty\left[\langle B(t) B'\rangle-\langle B\rangle \langle B'\rangle\right]e^{i\omega t} dt.
\end{equation}
Since the average is over the ground state, $S_{BB'}(\omega)=0$ for $\omega<0$.
If $B'=B^\dagger$, then $S_{BB^\dagger}(\omega)$ is positive and is known as the spectral density of $B$.

Another useful quantity is the static susceptibility $\chi_{BB'}$ defined as the change in the expectation value of $B$ under an infinitesimal variation of the Hamiltonian by $-\epsilon B'$:
\begin{equation}
\langle B\rangle_{H-\epsilon B'}=\langle B\rangle_{H}+\epsilon\chi_{BB'}+O(\epsilon^2).
\end{equation}
Here it is assumed that $B,B'$ are Hermitian, but one can formally extend $\chi_{BB'}$ to general operators by linearity. In linear response theory, it is shown that $\chi_{BB'}$  can be expressed through $S_{BB'}(\omega)$. At $T=0$ the formula looks as follows:
\begin{equation}
\chi_{BB'}=\int_0^\infty \frac{S_{BB'}(\omega)+S_{B'B}(\omega)}{2\omega}\frac{d\omega}{\pi}.
\end{equation}
The static susceptibility is not always well-defined because of a possible non-integrable singularity at $\omega=0$. Assuming it exists for $B=B'=Q(f)-\langle Q(f)\rangle$ and using the Cauchy-Schwarz inequality, we get an estimate:
\begin{equation}
 \begin{split}
      \langle B^\dagger B+B B^\dagger \rangle =    2\langle B(t) B \rangle \Big|_{t=0} = \int_0^\infty  S_{BB}(\omega)\frac{d\omega}{\pi}\\
     \le \sqrt{\int_0^\infty \omega S_{BB}(\omega) \frac{d\omega}{\pi} \int_0^\infty \frac{ S_{BB}(\omega)}{\omega}\frac{d\omega}{\pi} }=\sqrt{ |\langle [B,[H,B]]\rangle| } \sqrt{\chi_{BB}}.
 \end{split}
 \end{equation}
 
The quantity $\chi_{BB}$ depends on the function $f$ and thus also on $R$, but with the choice of $f$ described at the end section \ref{sec: Q estimates} it can be upper-bounded by a number of order $R^d$ times an $R$-independent quantity with a simple thermodynamic interpretation:
 \begin{align}  \label{eq: q sus estimate}
     \chi_{BB}=\sum_{{\bf n},{\bf m}} f({\bf n}) f({\bf m}) \chi_{Q_{\bf n} Q_{\bf m}} \le  C_d R^d \sum_{\bf n} \chi_{Q_0 Q_{\bf n}} +o(R^d)= C_d R^d \frac{\partial \langle Q_0\rangle } {\partial \mu}+o(R^d).
 \end{align}
 Here  $\mu$ is the chemical potential. The non-negative quantity $\partial \langle Q_0\rangle/\partial\mu$ is the derivative of the charge density with respect to the chemical potential, also known as charge compressibility. We assumed here that it is finite. 
 
 Combining (\ref{eq: q double com estimate}, \ref{eq: clustering estimate},\ref{eq: uncertainty relation}, \ref{eq: q sus estimate}) we find
 
 \begin{align}\label{eq: PS charge}
    | \langle [Q,A] \rangle|^2 \leq  \sqrt{\frac{\partial \langle Q_0\rangle } {\partial \mu}}\times  O(R^{d-2-\delta}).
 \end{align}
 Therefore if charge compressibility is finite and the ground state is clustering, charge symmetry cannot be broken for  $d\le2$. Note that in the absence of dipole symmetry one can only show that the charge symmetry cannot be broken for $d\leq 1$. 
 
We cannot use the same method to exclude dipole symmetry breaking in $d=1$. The main issue is that the analog of susceptibility for the dipole symmetry is 
 \begin{align}
     \chi_{D^i(f)D^i(f)} = \sum_{{\bf n},{\bf m}} n^im^if({\bf n})f({\bf m}) \chi_{Q_{\bf n},Q_{\bf m}},
 \end{align}
Even if we assume that $\chi_{Q_{\bf n},Q_{\bf m}}$ is essentially non-zero only if ${\bf n},{\bf m}$ are close to each other, the sum behaves as $O(R^{d+2})$. This is not enough to exclude symmetry breaking in $d=1$. In fact, the Lifshitz model shows that dipole symmetry can be spontaneously broken in $d=1$ for $T=0$.

\subsection{The dipolar Bose-Hubbard model}

The above theorems apply, in particular, to lattice models of fermions with a dipole symmetry. They cannot be directly applied to the dipolar Bose-Hubbard model because the on-site Hilbert space is infinite-dimensional, and the operators $Q_{\bf n},H_{\bf n}$ are unbounded. However, after a  slight modification, the proofs work for this model too, if local charge fluctuations in the relevant equilibrium state are not too wild. Indeed, we can write
\begin{equation}
H_{\bf n}=H_{\bf n}^{(1)}+H_{\bf n}^{(2)},\quad 
\end{equation}
where $H_{\bf n}^{(1)}$ is the dipole hopping term and $H_{\bf n}^{(2)}$ is a function of only $Q_{\bf m}$ on the neighboring sites. Then
\begin{equation}
[Q_{\bf m},H_{\bf k}]=[Q_{\bf m},H_{\bf k}^{(1)}]=\epsilon_{{\bf m}-{\bf k}} H^{(1)}_{\bf k},
\end{equation}
where $\epsilon_{{\bf m}-{\bf k}}$ is $\pm 2$ if ${\bf m}={\bf k}$, $\pm 1$ if ${\bf m},{\bf k}$ are nearest neighbors, and $0$ otherwise. Therefore the estimate (\ref{eq: double comm estimate}) can be replaced with
\begin{equation}
\left|\langle[Q_{\bf n}, [Q_{\bf m},H_{\bf k}]]\rangle\right|\leq C_d \left|\langle H_0\rangle\right|,
\end{equation}
where $C_d$ depends only on $d$. Further, it is easy to see that
\begin{equation}
\left|\langle H_0\rangle\right|\leq C'_d|t| \left|\langle Q_{\bf n}^2\rangle\right|,
\end{equation}
where $C'_d$ is another $d$-dependent constant. Thus if we assume that fluctuations of the charge density have a finite standard deviation, we can still bound the double commutator by a (state-dependent) constant. Note also that while the local operator $A_0$ need not be bounded (for example, it could be $b_0$), the clustering property implies that it has a finite standard deviation. Then we still get the estimates (\ref{eq: HMW charge}), (\ref{eq: HMW dipole}) and (\ref{eq: PS charge}). 

Similar arguments can be used for more general systems with unbounded $Q_{\bf n}$ and $H_{\bf n}$. As long as the fluctuations of the constituents are not too wild one can bound the expectation value on the left-hand side of (\ref{eq: double comm estimate}).

\section{Continuum theories}\label{sec: continuum}
\subsection{HMW-type theorems for continuum theories}\label{sec: HMWcont}

For continuum systems, the momentum density operator $p_j({\bf x})$ is an obvious candidate for an order parameter for the dipole symmetry. Indeed, if we choose the dipole charge density to be $d_j({\bf x}) = x_j \rho({\bf x})$, where $\rho({\bf x})$ is the charge density, we find
\begin{align}\label{eq: D p com}
    [D_j, p_k({\bf x})]= i\delta_{jk}\rho({\bf x}).
\end{align}
Thus, as long as the average charge density is not zero, the dipole symmetry appears to be broken. This statement does not depend on either temperature of dimensionality or space and suggests that the theorem of section \ref{sec:dipole} cannot be extended to continuum models with a dipole symmetry. 

In this section using a specific Hamiltonian as an example, we will show that the arguments of previous sections can be generalized without essential change to the continuum case. We discuss how the apparent contradiction is avoided. 

Consider a bosonic field $\psi$ with canonical commutation relations
\begin{align}
    [\psi({\bf x}),\psi^\dagger({\bf y})] &= \delta({\bf x}-{\bf y}),\\
    [\psi({\bf x}),\psi({\bf y})]&= [\psi^\dagger({\bf x}),\psi^\dagger({\bf y})]=0,
\end{align}
and the Hamiltonian
\begin{align}\label{eq: cont model Hamiltonian}
H&=  H_{\rm kin}+ H_{\rm pot},
\end{align}
where the kinetic energy is
\begin{align*}
    H_{\rm kin} &= \int d^d x \Big[\frac {K_1}2  \Big|\partial_j \psi \partial_k\psi- \psi \partial_j\partial_k\psi\Big|^2+\frac {K_2}2  (\partial_i \psi^\dagger \partial_j\psi^\dagger- \psi^\dagger \partial_j\partial_k\psi^\dagger) (\partial_j \psi \partial_k\psi- \psi \partial_j\partial_k\psi)\Big],
\end{align*}
 and the potential energy depends only on the density~$\rho = \psi^\dagger \psi$ but is arbitrary otherwise and can include long-range interactions. This Hamiltonian is invariant under charge and dipole transformations  $\psi\rightarrow \exp(i a+ ib_j x_j)\psi$. Similar Hamiltonians were  studied in \cite{FractonSuperfluid, LargeNfractons}.
 
 The charge conservation equation  has the form
 \begin{align}
    \dot\rho &= \partial_j\partial_k j_{jk},
\end{align}
where we defined
\begin{align}
    j_{jk}&=i\frac {K_1}2\,\big(\partial_j\psi^\dagger \partial_k\psi^\dagger-\psi^\dagger \partial_j\partial_k \psi^\dagger\big)\psi^2 +i\frac {K_2}2\,\delta_{jk} \big(\partial_l \psi^\dagger \partial_l\psi^\dagger-\psi^\dagger \partial^2 \psi^\dagger\big)\psi^2+h.c.
\end{align}
The commutator of the charge density with the current is
\begin{align}\label{eq: rhojcomm}
    [\rho({\bf x}),j_{jk}({\bf y})] = -iK_1 \partial_j \partial_k\delta({\bf x}-{\bf y}) {\psi^\dagger}^2 \psi^2({\bf y}) - iK_2 \delta_{jk}\partial^2\delta({\bf x}-{\bf y}) {\psi^\dagger}^2 \psi^2({\bf y}).
\end{align}

Using this commutation relation, we can do manipulations analogous to (\ref{eq: Q double com}):

\begin{align}\label{eq: cont QQ com est}
\begin{split}
    \int d^dxd^dy \, f({\bf x})f({\bf y}) \langle [\rho({\bf x}),[\rho({\bf y}),H]] \rangle =   i \int d^dxd^dy \, f({\bf x})f({\bf y}) \langle [\rho({\bf x}),\partial_j\partial_k j_{jk}({\bf y})] \rangle \\=  \int d^dxd^dy \,  \Big(K_1\partial_j\partial_k f(x) \partial_j\partial_k f({\bf y})+K_2 \partial^2f(x) \partial ^2f({\bf y})\Big)  \langle {\psi^\dagger}^2 \psi^2({\bf y})\rangle.
\end{split}
\end{align}

We can again apply the Bogoliubov inequality (\ref{eq: Bog ineq}) to $A=A_R=\dfrac 1 {{\rm Vol}\, B_R^d}\int_{B_R^d} d^dx\, \Big(a({\bf x})-\langle a({\bf x}) \rangle\Big)$ and $B=D_j(g)=\int d^dx \, x_j g({\bf x}) \rho({\bf x})$, where $a({\bf x})$ is order parameter, $B_R^d$ is a $d$-dimensional ball of radius $R$ centered at $0$, and the function $g({\bf x})$ is chosen to be $1$ inside of this ball and interpolates to 0 outside of $|{\bf x}|>2R$ so that second derivatives of $f({\bf x}) = x_j g({\bf x})$ are bounded by $O(\frac 1 R )$. We find
\begin{align}
    \Big| \langle[D,a(0)]\rangle \Big |^2 \le  \langle [\rho(f),[H,\rho(f)]] \rangle  \langle A_R^\dagger A_R +A_R A_R^\dagger\rangle
\end{align}
We can use (\ref{eq: cont QQ com est}) to show that
\begin{align}\label{eq: rhorhoHcomm}
    \langle [\rho(f),[H,\rho(f)]] \rangle  = O(R^{d-2}).
\end{align} 
Using the estimate (\ref{eq: clustering estimate}), we find
\begin{align}
    \Big| \langle[D,a(0)]\rangle \Big |^2 \le O(R^{d-2-\delta}).
\end{align}
Similarly, one can generalize the arguments  for the charge symmetry at zero or nonzero temperature leading to the same results as for lattice systems.

There are several ways in which the HMW theorem for the dipole symmetry and the commutation relation (\ref{eq: D p com}) can be reconciled. Most obviously, the momentum density operator may fail to  satisfy the clustering property. For example, take the Hamiltonian (\ref{eq: cont model Hamiltonian}) and choose the potential term to have a minimum at $|\rho|=\rho_0$. Then we can ignore the fluctuation of the modulus of $\psi$ and write approximately $\psi=\rho_0 \exp(i\phi)$. The effective action for $\phi$ is 
\begin{align}
    \mathcal S_{\rm eff} = \frac k 2\int d^dxdt \, \Big[\dot\phi^2-C_1(\partial^2\phi)^2\Big].
\end{align}
This is the Lifshitz model discussed in Section \ref{sec: Lifshitz}.
The charge density operator in this approximation becomes $k\dot\phi$. Thus in a state with a constant charge density we have $\phi=\mu t+\eta$, where $\mu$ is the chemical potential and $\eta$ is the fluctuating field with zero expectation value. The momentum density operator is $p_j = k\dot\phi\partial_j\phi $. To leading order in $\eta$ we get $p_j=k\mu\partial_j\eta$. As discussed in Section \ref{sec: Lifshitz}, the two-point function of  $\partial_j\eta$ grows with distance if $T>0$ and $d=1,2$. Therefore the same is true for $p_j$ provided $\mu\neq 0$. Note that if the two-point function of a local operator grows with distance, fluctuations of its spatial average increase as the averaging region becomes larger. This is why it is natural to require an operator serving as an order parameter to satisfy clustering.

Another possible loophole in the HMW theorem is that the spatially-smeared momentum density operator may fail to be bounded, and thus the expression $\langle A_R^\dagger A_R+A_R A_R^\dagger\rangle$ may be infinite. For this to happen, $\langle p_j({\bf x})p_k({\bf y})\rangle$ must have a non-integrable singularity at ${\bf x}={\bf y}$. However, this is a short-distance issue that can be fixed by replacing the momentum density operator with a suitably regularized expression. 

\subsection{Inertial mass density}

In Section \ref{sec: intro} we motivated the dipole symmetry by arguing that it emerges in the limit of infinite particle mass. For a large class of continuous systems with dipole symmetry, one can be more precise and prove that they have infinite inertial mass density.

It is well-known that the Bogolyubov inequality can be strengthened as follows:
\begin{equation}
\left|\langle [A,B]\rangle\right|^2\leq  \chi_{AA^\dagger}\,\langle [B,[H,B^\dagger]]\rangle .
\end{equation}
This is simply the Cauchy-Schwarz inequality for the Hermitian inner product $(A,C)=\chi_{AC^\dagger}$ applied to $C=i[H,B]$. Let us apply it to Hermitian operators $A=\dfrac 1 {{\rm vol}\, B_R^d} \int_{B^d_R} p_j({\bf x}) d^dx$ and $B=D_j(g)$, as above. Assuming that translational symmetry is unbroken, we get
\begin{equation}
\langle \rho(0)\rangle^2\leq \chi_{AA}\, \langle [\rho(f),[H,\rho(f)]]\rangle .
\end{equation}
Here $f(\bx)=x_j g(\bx)$ as before. For the model considered in Section \ref{sec: HMWcont}, or for any continuum theory with a dipole symmetry where the commutator of $\rho$ and $j_{jk}$ has the form similar to (\ref{eq: rhojcomm}), we have the estimate (\ref{eq: rhorhoHcomm}). Hence
\begin{equation}\label{eq: chiAAestimate}
\chi_{AA}\geq \langle \rho(0)\rangle ^2\times O(R^{2-d}). 
\end{equation}

To relate $\chi_{AA}$ to inertial mass density, recall that whenever  momentum is conserved, one can define a generalized Gibbs ensemble by replacing the Hamiltonian $H$ with $H-{\bf v}\cdot {\bf P}$, where ${\bf P}$ is the total momentum. The chemical potential for ${\bf P}$ is the velocity ${\bf v}$. It is natural to define the inertial mass density as the tensor
\begin{equation}
m_{jk}=\frac{\partial\langle p_j(0)\rangle}{\partial v_k}=\int \chi_{p_j(0)p_k({\bf x})} d^dx.   
\end{equation}
If $m_{jk}$ is finite, then $\chi_{AA}\leq m_{jj} \times O(R^{-d})$, which contradicts the inequality (\ref{eq: chiAAestimate}) for any nonzero $\langle\rho(0)\rangle$. 

\section{Discussion}\label{sec: discussion}

In this paper, we generalized the Hohenberg-Mermin-Wagner theorem to the dipole symmetry. We showed that for systems with finite-range interactions and clustering, dipole symmetry cannot be broken if $d=1,2$ and $T>0$. Thus a system of fermions with a microscopic dipole symmetry cannot flow to a Fermi liquid or any phase with a pronounced Fermi surface. Any phase of such a system must be rather exotic.

We also showed that the mere presence of a dipole symmetry allows one to strengthen the conclusions of the HMW and Pitaevskii-Stringari theorems as applied to charge symmetry. Namely, with the same assumptions, the charge symmetry cannot be broken if $d\leq 4, T>0$ and for $d\leq 2, T=0$ and  finite compressibility. 

Systems with a dipole symmetry  provide a counter-example to the widespread belief that spontaneous breaking of charge symmetry implies superconductivity (i.e. infinite d.c. conductivity). Indeed, for any system with a dipole symmetry the current operator vanishes at zero wave-vector and thus  conductivity vanishes at all frequencies. On the other hand, the example of the $d=3$ Lifshitz model at $T=0$ shows that systems with a dipole symmetry can break charge symmetry.

Charge symmetry and continuous translational symmetry are alike in some respects. For example, HMW-type theorems for them are essentially the same. Superficially, dipole symmetry should be similar to a continuous rotational symmetry: the dipole moment is an integral of charge density times a linear function of coordinates, just like the angular momentum is the integral of momentum density times a linear function of coordinates. Therefore it might seem surprising that rotational symmetry can be spontaneously broken in two dimensions for $T>0$ \cite{Mermin-classical}, while dipole symmetry cannot. The difference arises from the different group structure of the full symmetry group and the corresponding  constraints on the local conservation laws. Dipole symmetry requires the charge current to be a total derivative: $j_j=-\partial_k j_{jk}$. In contrast, rotational symmetry does not require the momentum current (i.e. the stress tensor) to be a total derivative, it only requires it to be symmetric. Consequently, the bounds resulting from the Bogolyubov inequality do not preclude the spontaneous breaking of rotational symmetry in two dimensions. This difference is also responsible for the different structure of the effective action for the Goldstone bosons.

Finally, it would be interesting to consider similar issues for systems of fermions in a magnetic field projected to the lowest Landau level. They have an infinite-dimensional GMP symmetry \cite{GMP} which contains a 1d dipole symmetry.

\bibliographystyle{apsrev4-1}
\bibliography{bib}

\begin{thebibliography}{28}%
\makeatletter
\providecommand \@ifxundefined [1]{%
 \@ifx{#1\undefined}
}%
\providecommand \@ifnum [1]{%
 \ifnum #1\expandafter \@firstoftwo
 \else \expandafter \@secondoftwo
 \fi
}%
\providecommand \@ifx [1]{%
 \ifx #1\expandafter \@firstoftwo
 \else \expandafter \@secondoftwo
 \fi
}%
\providecommand \natexlab [1]{#1}%
\providecommand \enquote  [1]{``#1''}%
\providecommand \bibnamefont  [1]{#1}%
\providecommand \bibfnamefont [1]{#1}%
\providecommand \citenamefont [1]{#1}%
\providecommand \href@noop [0]{\@secondoftwo}%
\providecommand \href [0]{\begingroup \@sanitize@url \@href}%
\providecommand \@href[1]{\@@startlink{#1}\@@href}%
\providecommand \@@href[1]{\endgroup#1\@@endlink}%
\providecommand \@sanitize@url [0]{\catcode `\\12\catcode `\$12\catcode
  `\&12\catcode `\#12\catcode `\^12\catcode `\_12\catcode `\%12\relax}%
\providecommand \@@startlink[1]{}%
\providecommand \@@endlink[0]{}%
\providecommand \url  [0]{\begingroup\@sanitize@url \@url }%
\providecommand \@url [1]{\endgroup\@href {#1}{\urlprefix }}%
\providecommand \urlprefix  [0]{URL }%
\providecommand \Eprint [0]{\href }%
\providecommand \doibase [0]{http://dx.doi.org/}%
\providecommand \selectlanguage [0]{\@gobble}%
\providecommand \bibinfo  [0]{\@secondoftwo}%
\providecommand \bibfield  [0]{\@secondoftwo}%
\providecommand \translation [1]{[#1]}%
\providecommand \BibitemOpen [0]{}%
\providecommand \bibitemStop [0]{}%
\providecommand \bibitemNoStop [0]{.\EOS\space}%
\providecommand \EOS [0]{\spacefactor3000\relax}%
\providecommand \BibitemShut  [1]{\csname bibitem#1\endcsname}%
\let\auto@bib@innerbib\@empty
%</preamble>
\bibitem [{\citenamefont {Peierls}(1934)}]{PeierlsG}%
  \BibitemOpen
  \bibfield  {author} {\bibinfo {author} {\bibfnamefont {R.~E.}\ \bibnamefont
  {Peierls}},\ }\bibfield  {title} {\enquote {\bibinfo {title} {Bemerkungen
  \"uber umwandlungstemperaturen},}\ }\href@noop {} {\bibfield  {journal}
  {\bibinfo  {journal} {Helv. Phys. Acta.}\ }\textbf {\bibinfo {volume} {7}},\
  \bibinfo {pages} {81} (\bibinfo {year} {1934})}\BibitemShut {NoStop}%
\bibitem [{\citenamefont {Peierls}(1935)}]{PeierlsF}%
  \BibitemOpen
  \bibfield  {author} {\bibinfo {author} {\bibfnamefont {R.~E.}\ \bibnamefont
  {Peierls}},\ }\bibfield  {title} {\enquote {\bibinfo {title} {Quelques
  propri{\'e}t{\'e}s typiques des corps solides},}\ }\href@noop {} {\bibfield
  {journal} {\bibinfo  {journal} {Annales de l'Institut Henri Poincar{\'e}}\
  }\textbf {\bibinfo {volume} {5}},\ \bibinfo {pages} {177} (\bibinfo {year}
  {1935})}\BibitemShut {NoStop}%
\bibitem [{\citenamefont {Landau}(1937)}]{Landau}%
  \BibitemOpen
  \bibfield  {author} {\bibinfo {author} {\bibfnamefont {L.~D.}\ \bibnamefont
  {Landau}},\ }\bibfield  {title} {\enquote {\bibinfo {title} {{On the theory
  of phase transitions, II}},}\ }\href@noop {} {\bibfield  {journal} {\bibinfo
  {journal} {Zh. Eksp. Teor. Fiz.}\ }\textbf {\bibinfo {volume} {7}},\ \bibinfo
  {pages} {627} (\bibinfo {year} {1937})}\BibitemShut {NoStop}%
\bibitem [{\citenamefont {Hohenberg}(1967)}]{Hohenberg}%
  \BibitemOpen
  \bibfield  {author} {\bibinfo {author} {\bibfnamefont {P.~C.}\ \bibnamefont
  {Hohenberg}},\ }\bibfield  {title} {\enquote {\bibinfo {title} {{Existence of
  Long-Range Order in One and Two Dimensions}},}\ }\href {\doibase
  10.1103/PhysRev.158.383} {\bibfield  {journal} {\bibinfo  {journal} {Phys.
  Rev.}\ }\textbf {\bibinfo {volume} {158}},\ \bibinfo {pages} {383} (\bibinfo
  {year} {1967})}\BibitemShut {NoStop}%
\bibitem [{\citenamefont {Mermin}\ and\ \citenamefont {Wagner}(1966)}]{MW}%
  \BibitemOpen
  \bibfield  {author} {\bibinfo {author} {\bibfnamefont {N.~D.}\ \bibnamefont
  {Mermin}}\ and\ \bibinfo {author} {\bibfnamefont {H.}~\bibnamefont
  {Wagner}},\ }\bibfield  {title} {\enquote {\bibinfo {title} {{Absence of
  Ferromagnetism or Antiferromagnetism in One- or Two-Dimensional Isotropic
  Heisenberg Models}},}\ }\href {\doibase 10.1103/PhysRevLett.17.1133}
  {\bibfield  {journal} {\bibinfo  {journal} {Phys. Rev. Lett.}\ }\textbf
  {\bibinfo {volume} {17}},\ \bibinfo {pages} {1133} (\bibinfo {year}
  {1966})}\BibitemShut {NoStop}%
\bibitem [{\citenamefont {Pitaevskii}\ and\ \citenamefont
  {Stringari}(1991)}]{Pitaevskii1991}%
  \BibitemOpen
  \bibfield  {author} {\bibinfo {author} {\bibfnamefont {L.}~\bibnamefont
  {Pitaevskii}}\ and\ \bibinfo {author} {\bibfnamefont {S.}~\bibnamefont
  {Stringari}},\ }\bibfield  {title} {\enquote {\bibinfo {title} {Uncertainty
  principle, quantum fluctuations, and broken symmetries},}\ }\href {\doibase
  10.1007/BF00682193} {\bibfield  {journal} {\bibinfo  {journal} {Journal of
  Low Temperature Physics}\ }\textbf {\bibinfo {volume} {85}},\ \bibinfo
  {pages} {377} (\bibinfo {year} {1991})}\BibitemShut {NoStop}%
\bibitem [{\citenamefont {{Berezinski{\v{i}}}}(1971)}]{Berezinski1}%
  \BibitemOpen
  \bibfield  {author} {\bibinfo {author} {\bibfnamefont {V.~L.}\ \bibnamefont
  {{Berezinski{\v{i}}}}},\ }\bibfield  {title} {\enquote {\bibinfo {title}
  {{Destruction of Long-range Order in One-dimensional and Two-dimensional
  Systems having a Continuous Symmetry Group I. Classical Systems}},}\
  }\href@noop {} {\bibfield  {journal} {\bibinfo  {journal} {Zh. Eksp. Teor.
  Fiz.}\ }\textbf {\bibinfo {volume} {59}},\ \bibinfo {pages} {907} (\bibinfo
  {year} {1971})}\BibitemShut {NoStop}%
\bibitem [{\citenamefont {{Berezinski{\v{i}}}}(1972)}]{Berezinski2}%
  \BibitemOpen
  \bibfield  {author} {\bibinfo {author} {\bibfnamefont {V.~L.}\ \bibnamefont
  {{Berezinski{\v{i}}}}},\ }\bibfield  {title} {\enquote {\bibinfo {title}
  {{Destruction of Long-range Order in One-dimensional and Two-dimensional
  Systems Possessing a Continuous Symmetry Group. II. Quantum Systems}},}\
  }\href@noop {} {\bibfield  {journal} {\bibinfo  {journal} {Zh. Eksp. Teor.
  Fiz.}\ }\textbf {\bibinfo {volume} {61}},\ \bibinfo {pages} {1144} (\bibinfo
  {year} {1972})}\BibitemShut {NoStop}%
\bibitem [{\citenamefont {Kosterlitz}\ and\ \citenamefont
  {Thouless}(1973)}]{Kosterlitz}%
  \BibitemOpen
  \bibfield  {author} {\bibinfo {author} {\bibfnamefont {J.~M.}\ \bibnamefont
  {Kosterlitz}}\ and\ \bibinfo {author} {\bibfnamefont {D.~J.}\ \bibnamefont
  {Thouless}},\ }\bibfield  {title} {\enquote {\bibinfo {title} {{Ordering,
  metastability and phase transitions in two-dimensional systems}},}\ }\href
  {\doibase 10.1088/0022-3719/6/7/010} {\bibfield  {journal} {\bibinfo
  {journal} {Journal of Physics C: Solid State Physics}\ }\textbf {\bibinfo
  {volume} {6}},\ \bibinfo {pages} {1181} (\bibinfo {year} {1973})}\BibitemShut
  {NoStop}%
\bibitem [{\citenamefont {{Lake}}\ \emph {et~al.}()\citenamefont {{Lake}},
  \citenamefont {{Hermele}},\ and\ \citenamefont {{Senthil}}}]{lakesenthil}%
  \BibitemOpen
  \bibfield  {author} {\bibinfo {author} {\bibfnamefont {E.}~\bibnamefont
  {{Lake}}}, \bibinfo {author} {\bibfnamefont {M.}~\bibnamefont {{Hermele}}}, \
  and\ \bibinfo {author} {\bibfnamefont {T.}~\bibnamefont {{Senthil}}},\
  }\bibfield  {title} {\enquote {\bibinfo {title} {{The dipolar Bose-Hubbard
  model}},}\ }\href@noop {} {\bibfield  {journal} {\bibinfo  {journal} {arXiv
  e-prints}\ }}\Eprint {http://arxiv.org/abs/2201.04132} {arXiv:2201.04132}
  \BibitemShut {NoStop}%
\bibitem [{\citenamefont {Haah}(2011)}]{Haah}%
  \BibitemOpen
  \bibfield  {author} {\bibinfo {author} {\bibfnamefont {J.}~\bibnamefont
  {Haah}},\ }\bibfield  {title} {\enquote {\bibinfo {title} {Local stabilizer
  codes in three dimensions without string logical operators},}\ }\href
  {\doibase 10.1103/PhysRevA.83.042330} {\bibfield  {journal} {\bibinfo
  {journal} {Phys. Rev. A}\ }\textbf {\bibinfo {volume} {83}},\ \bibinfo
  {pages} {042330} (\bibinfo {year} {2011})}\BibitemShut {NoStop}%
\bibitem [{\citenamefont {Vijay}\ \emph {et~al.}(2015)\citenamefont {Vijay},
  \citenamefont {Haah},\ and\ \citenamefont {Fu}}]{Fu}%
  \BibitemOpen
  \bibfield  {author} {\bibinfo {author} {\bibfnamefont {S.}~\bibnamefont
  {Vijay}}, \bibinfo {author} {\bibfnamefont {J.}~\bibnamefont {Haah}}, \ and\
  \bibinfo {author} {\bibfnamefont {L.}~\bibnamefont {Fu}},\ }\bibfield
  {title} {\enquote {\bibinfo {title} {A new kind of topological quantum order:
  A dimensional hierarchy of quasiparticles built from stationary
  excitations},}\ }\href {\doibase 10.1103/PhysRevB.92.235136} {\bibfield
  {journal} {\bibinfo  {journal} {Phys. Rev. B}\ }\textbf {\bibinfo {volume}
  {92}},\ \bibinfo {pages} {235136} (\bibinfo {year} {2015})}\BibitemShut
  {NoStop}%
\bibitem [{\citenamefont {Nandkishore}\ and\ \citenamefont
  {Hermele}(2019)}]{FractonReview}%
  \BibitemOpen
  \bibfield  {author} {\bibinfo {author} {\bibfnamefont {R.~M.}\ \bibnamefont
  {Nandkishore}}\ and\ \bibinfo {author} {\bibfnamefont {M.}~\bibnamefont
  {Hermele}},\ }\bibfield  {title} {\enquote {\bibinfo {title} {Fractons},}\
  }\href {\doibase 10.1146/annurev-conmatphys-031218-013604} {\bibfield
  {journal} {\bibinfo  {journal} {Annual Review of Condensed Matter Physics}\
  }\textbf {\bibinfo {volume} {10}},\ \bibinfo {pages} {295} (\bibinfo {year}
  {2019})}\BibitemShut {NoStop}%
\bibitem [{\citenamefont {Pretko}(2017)}]{Pretko}%
  \BibitemOpen
  \bibfield  {author} {\bibinfo {author} {\bibfnamefont {M.}~\bibnamefont
  {Pretko}},\ }\bibfield  {title} {\enquote {\bibinfo {title} {Subdimensional
  particle structure of higher rank $u(1)$ spin liquids},}\ }\href {\doibase
  10.1103/PhysRevB.95.115139} {\bibfield  {journal} {\bibinfo  {journal} {Phys.
  Rev. B}\ }\textbf {\bibinfo {volume} {95}},\ \bibinfo {pages} {115139}
  (\bibinfo {year} {2017})}\BibitemShut {NoStop}%
\bibitem [{\citenamefont {Pretko}(2018)}]{Pretko2}%
  \BibitemOpen
  \bibfield  {author} {\bibinfo {author} {\bibfnamefont {M.}~\bibnamefont
  {Pretko}},\ }\bibfield  {title} {\enquote {\bibinfo {title} {The fracton
  gauge principle},}\ }\href {\doibase 10.1103/PhysRevB.98.115134} {\bibfield
  {journal} {\bibinfo  {journal} {Phys. Rev. B}\ }\textbf {\bibinfo {volume}
  {98}},\ \bibinfo {pages} {115134} (\bibinfo {year} {2018})}\BibitemShut
  {NoStop}%
\bibitem [{\citenamefont {Gorantla}\ \emph {et~al.}(2022)\citenamefont
  {Gorantla}, \citenamefont {Lam}, \citenamefont {Seiberg},\ and\ \citenamefont
  {Shao}}]{Seiberg}%
  \BibitemOpen
  \bibfield  {author} {\bibinfo {author} {\bibfnamefont {P.}~\bibnamefont
  {Gorantla}}, \bibinfo {author} {\bibfnamefont {H.~T.}\ \bibnamefont {Lam}},
  \bibinfo {author} {\bibfnamefont {N.}~\bibnamefont {Seiberg}}, \ and\
  \bibinfo {author} {\bibfnamefont {S.-H.}\ \bibnamefont {Shao}},\ }\bibfield
  {title} {\enquote {\bibinfo {title} {{Global dipole symmetry, compact
  Lifshitz theory, tensor gauge theory, and fractons}},}\ }\href {\doibase
  10.1103/PhysRevB.106.045112} {\bibfield  {journal} {\bibinfo  {journal}
  {Phys. Rev. B}\ }\textbf {\bibinfo {volume} {106}},\ \bibinfo {pages}
  {045112} (\bibinfo {year} {2022})}\BibitemShut {NoStop}%
\bibitem [{\citenamefont {Abrikosov}\ \emph {et~al.}(1975)\citenamefont
  {Abrikosov}, \citenamefont {Dzyaloshinskii}, \citenamefont {Gorkov},\ and\
  \citenamefont {Silverman}}]{AGD}%
  \BibitemOpen
  \bibfield  {author} {\bibinfo {author} {\bibfnamefont {A.~A.}\ \bibnamefont
  {Abrikosov}}, \bibinfo {author} {\bibfnamefont {I.}~\bibnamefont
  {Dzyaloshinskii}}, \bibinfo {author} {\bibfnamefont {L.~P.}\ \bibnamefont
  {Gorkov}}, \ and\ \bibinfo {author} {\bibfnamefont {R.~A.}\ \bibnamefont
  {Silverman}},\ }\href {https://cds.cern.ch/record/107441} {\emph {\bibinfo
  {title} {{Methods of quantum field theory in statistical physics}}}}\
  (\bibinfo  {publisher} {Dover},\ \bibinfo {address} {New York, NY},\ \bibinfo
  {year} {1975})\BibitemShut {NoStop}%
\bibitem [{\citenamefont {Yuan}\ \emph {et~al.}(2020)\citenamefont {Yuan},
  \citenamefont {Chen},\ and\ \citenamefont {Ye}}]{FractonSuperfluid}%
  \BibitemOpen
  \bibfield  {author} {\bibinfo {author} {\bibfnamefont {J.-K.}\ \bibnamefont
  {Yuan}}, \bibinfo {author} {\bibfnamefont {S.~A.}\ \bibnamefont {Chen}}, \
  and\ \bibinfo {author} {\bibfnamefont {P.}~\bibnamefont {Ye}},\ }\bibfield
  {title} {\enquote {\bibinfo {title} {Fractonic superfluids},}\ }\href
  {\doibase 10.1103/PhysRevResearch.2.023267} {\bibfield  {journal} {\bibinfo
  {journal} {Phys. Rev. Research}\ }\textbf {\bibinfo {volume} {2}},\ \bibinfo
  {pages} {023267} (\bibinfo {year} {2020})}\BibitemShut {NoStop}%
\bibitem [{\citenamefont {Stahl}\ \emph {et~al.}(2022)\citenamefont {Stahl},
  \citenamefont {Lake},\ and\ \citenamefont {Nandkishore}}]{Lake2}%
  \BibitemOpen
  \bibfield  {author} {\bibinfo {author} {\bibfnamefont {C.}~\bibnamefont
  {Stahl}}, \bibinfo {author} {\bibfnamefont {E.}~\bibnamefont {Lake}}, \ and\
  \bibinfo {author} {\bibfnamefont {R.}~\bibnamefont {Nandkishore}},\
  }\bibfield  {title} {\enquote {\bibinfo {title} {Spontaneous breaking of
  multipole symmetries},}\ }\href {\doibase 10.1103/PhysRevB.105.155107}
  {\bibfield  {journal} {\bibinfo  {journal} {Phys. Rev. B}\ }\textbf {\bibinfo
  {volume} {105}},\ \bibinfo {pages} {155107} (\bibinfo {year}
  {2022})}\BibitemShut {NoStop}%
\bibitem [{\citenamefont {Watanabe}\ and\ \citenamefont
  {Vishwanath}(2014)}]{Watanabe}%
  \BibitemOpen
  \bibfield  {author} {\bibinfo {author} {\bibfnamefont {H.}~\bibnamefont
  {Watanabe}}\ and\ \bibinfo {author} {\bibfnamefont {A.}~\bibnamefont
  {Vishwanath}},\ }\bibfield  {title} {\enquote {\bibinfo {title} {{Criterion
  for stability of Goldstone modes and Fermi liquid behavior in a metal with
  broken symmetry}},}\ }\href {\doibase 10.1073/pnas.1415592111} {\bibfield
  {journal} {\bibinfo  {journal} {Proceedings of the National Academy of
  Sciences}\ }\textbf {\bibinfo {volume} {111}},\ \bibinfo {pages} {16314}
  (\bibinfo {year} {2014})},\ \Eprint
  {http://arxiv.org/abs/https://www.pnas.org/doi/pdf/10.1073/pnas.1415592111}
  {https://www.pnas.org/doi/pdf/10.1073/pnas.1415592111} \BibitemShut {NoStop}%
\bibitem [{\citenamefont {Alberte}\ and\ \citenamefont
  {Nicolis}(2020)}]{Alberte}%
  \BibitemOpen
  \bibfield  {author} {\bibinfo {author} {\bibfnamefont {L.}~\bibnamefont
  {Alberte}}\ and\ \bibinfo {author} {\bibfnamefont {A.}~\bibnamefont
  {Nicolis}},\ }\bibfield  {title} {\enquote {\bibinfo {title} {{Spontaneously
  broken boosts and the Goldstone continuum}},}\ }\href {\doibase
  10.1007/JHEP07(2020)076} {\bibfield  {journal} {\bibinfo  {journal} {Journal
  of High Energy Physics}\ }\textbf {\bibinfo {volume} {2020}},\ \bibinfo
  {pages} {76} (\bibinfo {year} {2020})}\BibitemShut {NoStop}%
\bibitem [{\citenamefont {{Radzihovsky}}\ and\ \citenamefont
  {{Hermele}}(2020)}]{LeoMike}%
  \BibitemOpen
  \bibfield  {author} {\bibinfo {author} {\bibfnamefont {L.}~\bibnamefont
  {{Radzihovsky}}}\ and\ \bibinfo {author} {\bibfnamefont {M.}~\bibnamefont
  {{Hermele}}},\ }\bibfield  {title} {\enquote {\bibinfo {title} {{Fractons
  from Vector Gauge Theory}},}\ }\href {\doibase
  10.1103/PhysRevLett.124.050402} {\bibfield  {journal} {\bibinfo  {journal}
  {\prl}\ }\textbf {\bibinfo {volume} {124}},\ \bibinfo {eid} {050402}
  (\bibinfo {year} {2020})}\BibitemShut {NoStop}%
\bibitem [{\citenamefont {{Radzihovsky}}(2020)}]{Leo}%
  \BibitemOpen
  \bibfield  {author} {\bibinfo {author} {\bibfnamefont {L.}~\bibnamefont
  {{Radzihovsky}}},\ }\bibfield  {title} {\enquote {\bibinfo {title} {{Quantum
  Smectic Gauge Theory}},}\ }\href {\doibase 10.1103/PhysRevLett.125.267601}
  {\bibfield  {journal} {\bibinfo  {journal} {\prl}\ }\textbf {\bibinfo
  {volume} {125}},\ \bibinfo {eid} {267601} (\bibinfo {year}
  {2020})}\BibitemShut {NoStop}%
\bibitem [{\citenamefont {Martin}(1982)}]{Martin1982}%
  \BibitemOpen
  \bibfield  {author} {\bibinfo {author} {\bibfnamefont {P.~A.}\ \bibnamefont
  {Martin}},\ }\bibfield  {title} {\enquote {\bibinfo {title} {{A remark on the
  Goldstone theorem in statistical mechanics}},}\ }\href {\doibase
  10.1007/BF02890151} {\bibfield  {journal} {\bibinfo  {journal} {Il Nuovo
  Cimento B (1971-1996)}\ }\textbf {\bibinfo {volume} {68}},\ \bibinfo {pages}
  {302} (\bibinfo {year} {1982})}\BibitemShut {NoStop}%
\bibitem [{\citenamefont {Momoi}(1996)}]{Momoi1996}%
  \BibitemOpen
  \bibfield  {author} {\bibinfo {author} {\bibfnamefont {T.}~\bibnamefont
  {Momoi}},\ }\bibfield  {title} {\enquote {\bibinfo {title} {Quantum
  fluctuations in quantum lattice systems with continuous symmetry},}\ }\href
  {\doibase 10.1007/BF02175562} {\bibfield  {journal} {\bibinfo  {journal}
  {Journal of Statistical Physics}\ }\textbf {\bibinfo {volume} {85}},\
  \bibinfo {pages} {193} (\bibinfo {year} {1996})}\BibitemShut {NoStop}%
\bibitem [{\citenamefont {Jensen}\ and\ \citenamefont
  {Raz}()}]{LargeNfractons}%
  \BibitemOpen
  \bibfield  {author} {\bibinfo {author} {\bibfnamefont {K.}~\bibnamefont
  {Jensen}}\ and\ \bibinfo {author} {\bibfnamefont {A.}~\bibnamefont {Raz}},\
  }\bibfield  {title} {\enquote {\bibinfo {title} {Large ${N}$ fractons},}\
  }\href {https://arxiv.org/abs/2205.01132} {\ }\Eprint
  {http://arxiv.org/abs/2205.01132} {arXiv:2205.01132} \BibitemShut {NoStop}%
\bibitem [{\citenamefont {{Mermin}}(1967)}]{Mermin-classical}%
  \BibitemOpen
  \bibfield  {author} {\bibinfo {author} {\bibfnamefont {N.~D.}\ \bibnamefont
  {{Mermin}}},\ }\bibfield  {title} {\enquote {\bibinfo {title} {{Absence of
  Ordering in Certain Classical Systems}},}\ }\href {\doibase
  10.1063/1.1705316} {\bibfield  {journal} {\bibinfo  {journal} {Journal of
  Mathematical Physics}\ }\textbf {\bibinfo {volume} {8}},\ \bibinfo {pages}
  {1061} (\bibinfo {year} {1967})}\BibitemShut {NoStop}%
\bibitem [{\citenamefont {Girvin}\ \emph {et~al.}(1986)\citenamefont {Girvin},
  \citenamefont {MacDonald},\ and\ \citenamefont {Platzman}}]{GMP}%
  \BibitemOpen
  \bibfield  {author} {\bibinfo {author} {\bibfnamefont {S.~M.}\ \bibnamefont
  {Girvin}}, \bibinfo {author} {\bibfnamefont {A.~H.}\ \bibnamefont
  {MacDonald}}, \ and\ \bibinfo {author} {\bibfnamefont {P.~M.}\ \bibnamefont
  {Platzman}},\ }\bibfield  {title} {\enquote {\bibinfo {title} {Magneto-roton
  theory of collective excitations in the fractional quantum hall effect},}\
  }\href {\doibase 10.1103/PhysRevB.33.2481} {\bibfield  {journal} {\bibinfo
  {journal} {Phys. Rev. B}\ }\textbf {\bibinfo {volume} {33}},\ \bibinfo
  {pages} {2481} (\bibinfo {year} {1986})}\BibitemShut {NoStop}%
\end{thebibliography}%
\end{document}